\documentclass[preprint,aps,nofootinbib,showpacs,amsfonts,epsf]{revtex4}

\input{epsf.tex}

\newcommand{\E}{{\cal{E}}}
\newcommand{\s}{\sigma}

\renewcommand{\a}{\alpha}

\newcommand{\dfrac}[2]{\displaystyle\frac{#1}{#2}}
\newcommand{\be}{\begin{equation}}
\newcommand{\ee}{\end{equation}}
\newcommand{\bea}{\begin{eqnarray}}
\newcommand{\eea}{\end{eqnarray}}
\newcommand{\ba}{\begin{array}}
\newcommand{\ea}{\end{array}}
\def\J#1#2#3#4{{#1} {\bf #2}, #3 (#4)}
\def\PRD{Phys. Rev. D}
\def\PR{Phys. Rev.}
\def\PRL{Phys. Rev. Lett.}
\def\PTP{Prog. Theor. Phys.}
\def\LRR{Living Rev. Relativ.}
\def\APN{Ann. Phys. (NY)}

\def\MNRAS{Mon. Not. R. Astron. Soc.}
\def\JMP{J. Math. Phys.}
\def\CPAM{Comm. Pure Appl. Math.}

\def\CQG{Class. Quantum Grav.}

\def\PLA{Phys. Lett. A}

\begin{document}
\draft
\title{Metric for two equal Kerr black holes}

\author{V.~S.~Manko$^\dagger$ and E.~Ruiz$^\ddagger$ }
\address{$^\dagger$Departamento de F\'\i sica, Centro de Investigaci\'on y
de Estudios Avanzados del IPN, A.P. 14-740, 07000 Ciudad de
M\'exico, Mexico\\$^\ddagger$Instituto Universitario de F\'{i}sica
Fundamental y Matem\'aticas, Universidad de Salamanca, 37008
Salamanca, Spain}

\begin{abstract}
We show that the exact solution of Einstein's equations describing
a system of two aligned identical Kerr black holes separated by a
massless strut follows straightforwardly from the extended
2-soliton solution possessing equatorial symmetry, and we give its
concise analytic representation in terms of physical parameters.
\end{abstract}


\maketitle

\section{Introduction}

In our recent paper \cite{MRu} we have considered in detail a
vacuum specialization of the general 2-soliton electrovac metric
\cite{MMR} (henceforth referred to as the MMR solution) in
application to the description of the exterior geometry of neutron
stars \cite{Ste,PAp}. Although we gave in that paper three
different representations of the vacuum MMR solution, still we had
one more representation of the latter solution that had been
constructed by us some time ago for treating the two-body
configurations of spinning black holes of Kerr's type \cite{Ker};
however, we left its consideration for the future because of the
specific objectives of the paper \cite{MRu}. The appearance of a
preprint \cite{Cab} devoted to a system of corotating Kerr sources
yet motivates us to publish our results on the vacuum MMR solution
not earlier included into the paper \cite{MRu}, especially taking
into account that although the authors of the paper \cite{Cab}
solve correctly the axis condition for the binary system, they
still offer only a complicated form of the resulting metric, in
the absence of some important details of the derivation that might
be interesting to the reader. Our present work will aim therefore
at working out a concise representation of the 3-parameter
subfamily of the MMR solution describing a system of two equal
Kerr black holes kept apart from falling onto each other by a
massless strut \cite{Isr}, that would be alternative to the
representation obtained in Ref.~\cite{Cab}. To accomplish this
goal, we will first rewrite, using the procedure we have developed
in a series of papers devoted to the binary black-hole
configurations \cite{Man,MRR,MRR2,MRS}, the vacuum MMR metric in
terms of physical parameters by taking as a starting point the
axis data of the extended equatorially symmetric 2-soliton
solution in the form \cite{EMR}
\be e(z)=\frac{z^2-\bar b_1z+\bar b_2} {z^2+b_1z+b_2},
\label{axis} \ee
where $b_1$ and $b_2$ are two arbitrary complex constants, and a
bar over a symbol means complex conjugation. The particular
3-parameter case of two separated Kerr black holes will then arise
after imposing the axis condition in the general 4-parameter
metric.

The paper is organized as follows. In the next section we perform
a reparametrization of the data (\ref{axis}) in terms of the
quantities $M$, $a$, $\s$ and $R$ related, respectively, to the
masses of the sources, their angular momenta, the horizons'
half-lengths and the coordinate distance between the centers of
the sources. The reparametrized axis data is then used for writing
out the MMR solution in a new concise representation with the aid
of the general formulas of Ref.~\cite{MRu2}. In Sec.~III we solve
the axis condition for the MMR solution and analyze the resulting
3-parameter configuration of corotating Kerr black holes, thus
confirming some of the results of Ref.~\cite{Cab}. By expanding
the expression of the interaction force in inverse powers of $R$,
we show in particular that the leading spin-spin repulsion term
has precisely the same form as was given earlier by Dietz and
Hoenselaers \cite{DHo} through the analysis of two limiting cases
of spinning particles. In Sec.~IV we give the reparametrized form
of the extended 2-soliton metric suitable for treating the case of
two non-equal Kerr black holes. Sec.~V contains concluding
remarks.

\section{Yet another representation of the vacuum MMR solution}

We would like to recall that the extended vacuum soliton solutions
\cite{MRu2} constructed with the aid of Sibgatullin's integral
method \cite{Sib} are written in terms of the parameters $\a_n$
and $\beta_l$, the former parameters taking real values or forming
complex conjugate pairs (these determine the location of sources
on the symmetry axis), and the latter being roots of the
denominator in the axis data, hence taking arbitrary complex
values.

In the 2-soliton case with the additional equatorial symmetry we
have $\a_1=-\a_4$, $\a_2=-\a_3$, so that the $\a$'s can be
parametrized as
\be \a_1=\frac{1}{2}R+\sigma, \quad \a_2=\frac{1}{2}R-\sigma,
\quad \a_3=-\frac{1}{2}R+\sigma, \quad \a_4=-\frac{1}{2}R-\sigma,
\label{alfas} \ee
or, inversely,
\be R=\frac{1}{2}(\a_1+\a_2-\a_3-\a_4), \quad
\sigma=\frac{1}{2}(\a_1-\a_2)=\frac{1}{2}(\a_3-\a_4), \label{Rs}
\ee
where $R$ is the coordinate distance between the centers of black
holes, and $\sigma$ is the half-length of the horizon of each
black hole (see Fig.~1). Note that $\s$ in the above formulas
(\ref{alfas}) and (\ref{Rs}), as well as throughout this paper,
can also take on pure imaginary values, in which case the solution
would describe a pair of equal hyperextreme objects. However,
except for some special occurrences, below we will restrict our
analysis to the black-hole configurations only.

To identify the complex parameters $\beta_1$ and $\beta_2$, one
has to introduce explicitly the axis data --- the value of the
Ernst complex potential \cite{Ern} on the upper part of the
symmetry axis. In our case such data is given by formula
(\ref{axis}), and obviously can be cast into the equivalent form
\be e(z)=\frac{z^2-2(M+ia)z+c+id} {z^2+2(M-ia)z+c-id},
\label{axis2} \ee
involving four arbitrary real constants $M$, $a$, $c$ and $d$.
Since $\beta_1$ and $\beta_2$ are roots of the denominator on the
right-hand side of (\ref{axis2}), it is clear that these verify
the relation $\beta_1+\beta_2=-2(M-ia)$ and $\beta_1\beta_2=c-id$,
while the denominator itself can be formally written as
$(z-\beta_1)(z-\beta_2)$.

We must bear in mind that the parameters $\a_n$ in Sibgatullin's
method satisfy the equation
\be e(z)+\bar e(z)=0, \label{Sibga} \ee
which means that if we want to introduce these $\a_n$ into the
2-soliton solution as arbitrary parameters in the form
(\ref{alfas}), then we have to solve the equation
\be e(z)+\bar e(z)= \frac{2(z-\a_1)(z-\a_2)(z-\a_3)(z-\a_4)}
{(z-\beta_1)(z-\beta_2)(z-\bar\beta_1)(z-\bar\beta_2)} \label{ecu}
\ee
for the constants $c$ and $d$ by equating the coefficients at the
same powers of $z$ on both sides of (\ref{ecu}). A simple algebra
then yields
\be c=-\frac{1}{4}R^2+2M^2-2a^2-\s^2, \quad
d=\epsilon\sqrt{(R^2-4M^2+4a^2)(\s^2-M^2+a^2)}, \quad
\epsilon=\pm1, \label{cd} \ee
with which the axis data (\ref{axis2}) finally takes the form
\be e(z)=\frac{z^2-2(M+ia)z-{\textstyle\frac14}R^2
+2M^2-2a^2-\s^2+id} {z^2+2(M-ia)z-{\textstyle\frac14}R^2
+2M^2-2a^2-\s^2-id}, \label{axis3} \ee
where the constant quantity $d$ has been defined in (\ref{cd}).

Therefore, we have rewritten the axis data (\ref{axis2})
containing the parameters $M$, $a$, $c$ and $d$ in the equivalent
form (\ref{axis3}) involving the desired set of the parameters
$M$, $a$, $R$ and $\s$. It is worth noting that while the physical
meaning of the constants $R$ and $\s$ is transparent, the
interpretation of the parameters $M$ and $a$ can be revealed by
calculating the solution's total mass $M_T$ and total angular
momentum $J_T$ from (\ref{axis3}) with the help of the Fodor {\it
et al.} procedure \cite{FHP} for the evaluation of Geroch-Hansen
multipole moments \cite{Ger,Han}. Thus we get
\be M_T=2M, \quad J_T=4Ma-d, \label{MJ} \ee
whence it follows immediately that $M$ is half the total mass of
the configuration, whereas $a$ is the rotational parameter.
Observe that $M$ does not coincide exactly with the mass of each
black-hole constituent because the intermediate region
$\{\rho=0,|z|<\a_2\}$ in Fig.~1 may in principle carry some mass,
positive or negative.

Once the axis data is worked out, the corresponding potential $\E$
satisfying the Ernst equation \cite{Ern},
\be (\E+\bar\E)\Delta\E =2(\nabla\E)^2, \label{EE} \ee
can be obtained from the formula \cite{MRu2}
\be \E=\frac{E_+}{E_-}, \quad E_\pm=\left|
\begin{array}{ccccc}
1 & 1 & 1 & 1 & 1 \\ \vspace{0.15cm} \pm1 &
\dfrac{r_{1}}{\a_{1}-\beta_{1}} & \dfrac{r_{2}} {\a_{2}-\beta_{1}}
& \dfrac{r_{3}}{\a_{3}-\beta_{1}} &
\dfrac{r_{4}}{\a_{4}-\beta_{1}} \\ \vspace{0.15cm}
 \pm1 & \dfrac{r_{1}}{\a_{1}-\beta_{2}} & \dfrac{r_{2}}{\a_{2}-\beta_{2}}
& \dfrac{r_{3}}{\a_{3}-\beta_{2}} &
\dfrac{r_{4}}{\a_{4}-\beta_{2}}
\\ \vspace{0.15cm}
 0 & \dfrac{1}{\a_{1}-\bar{\beta}_{1}} &
\dfrac{1}{\a_{2}-\bar{\beta}_{1}} &
\dfrac{1}{\a_{3}-\bar{\beta}_{1}} &
\dfrac{1}{\a_{4}-\bar{\beta}_{1}} \\ \vspace{0.15cm}
 0 & \dfrac{1}{\a_{1}-\bar{\beta}_{2}} &
\dfrac{1}{\a_{2}-\bar{\beta}_{2}} &
\dfrac{1}{\a_{3}-\bar{\beta}_{2}} &
\dfrac{1}{\a_{4}-\bar{\beta}_{2}}
\end{array}
\right|, \label{EP} \ee
by just substituting the expressions of $\a$'s and $\beta$'s
determined by (\ref{alfas}) and (\ref{axis3}) into (\ref{EP}), and
taking into account that the functions $r_n$, which depend on the
coordinates $\rho$ and $z$, have the form
$r_n=\sqrt{\rho^2+(z-\a_n)^2}$.

In the Ernst formalism \cite{Ern}, the knowledge of the potential
$\E$ is sufficient for the construction of the corresponding
metric functions $f$, $\gamma$ and $\omega$ from the stationary
axisymmetric line element
\be d s^2=f^{-1}[e^{2\gamma}(d\rho^2+d z^2)+\rho^2 d\varphi^2]-f(d
t-\omega d\varphi)^2, \label{Papa} \ee
and the explicit expressions for these functions defined by the
potential (\ref{EP}) can be found in Ref.~\cite{MRu2} both in the
form of determinants and in the expanded form most suitable for
concrete computations and presentation of the results. Our own
evaluation of $\E$, $f$, $\gamma$ and $\omega$ for the axis data
(\ref{axis3}) yields the following final formulas:
\bea \E&=&\frac{A-B}{A+B}, \quad f=\frac{A\bar A-B\bar
B}{(A+B)(\bar A+\bar B)}, \quad e^{2\gamma}=\frac{A\bar A-B\bar
B}{K_0^2R_+R_-r_+r_-}, \nonumber\\
\omega&=&4a-\frac{2{\rm Im}[G(\bar A+\bar B)]}{A\bar A-B\bar B},
\nonumber\\ A&=&R^2(R_+-R_-)(r_+-r_-)
-4\s^2(R_+-r_+)(R_--r_-), \nonumber\\
B&=&2R\s[(R+2\s)(R_--r_+)-(R-2\s)(R_+-r_-)], \nonumber\\
G&=&-zB +R\s[2R(R_-r_--R_+r_+) +4\s(R_+R_--r_+r_-)\nonumber\\
&&-(R^2-4\s^2)(R_+-R_--r_++r_-)], \label{mfn} \eea
where
\bea R_\pm&=&\frac{-M(\pm2\s+R)+id}{2M^2+(R+2ia)(\pm\s+ia)}
\sqrt{\rho^2+\left(z+\frac{1}{2}R\pm\s\right)^2}, \nonumber\\
r_\pm&=&\frac{-M(\pm2\s-R)+id}{2M^2-(R-2ia)(\pm\s+ia)}
\sqrt{\rho^2+\left(z-\frac{1}{2}R\pm\s\right)^2}, \label{Rrn} \eea
and
\be K_0=\frac{4R^2\s^2(R^2-4\s^2)[(R^2+4a^2)(\s^2+a^2)
-4M(M^3+ad)]}{[M^2(R+2\s)^2+d^2][M^2(R-2\s)^2+d^2]}. \label{K0}
\ee

Eqs.~(\ref{mfn})-(\ref{K0}) and (\ref{cd}) fully determine the
desired representation of the 4-parameter vacuum MMR solution
which, as will be seen in the next section, is very suitable for
treating the case of two separated Kerr black holes. One can check
by direct calculation that on the upper part of the symmetry axis
$\{\rho=0,z>{\textstyle\frac{1}{2}}R+\sigma\}$ the potential $\E$
in (\ref{mfn}) reduces to the axis data (\ref{axis3}).

\section{Two identical Kerr black holes separated by a strut}

The MMR solution discussed in the previous section can be
interpreted as describing a pair of corotating Kerr black holes
after subjecting its parameters to the constraint
\be \omega=0 \quad \mbox{for} \quad \rho=0,\,\, |z|<\frac12R-\s,
\label{ac} \ee
which is known as the axis condition; this being satisfied,
converts the region $\{\rho=0, |z|<\frac12R-\s\}$ into a massless
conical singularity, a strut \cite{Isr}, which separates the two
black-hole constituents and prevents them from falling onto each
other. In this special case, the parameter $M$ becomes equal to
the Komar mass \cite{Kom} of each constituent exactly, while the
individual angular momentum $J$ of each black hole becomes equal
to $J_T/2$ because the strut does not make contribution into the
mass and angular momentum of the configuration.

On the symmetry axis, the metric function $\omega$ of the
2-soliton metric takes constant values generically \cite{TKi}, so
that from the condition (\ref{ac}) we get a (complicated)
algebraic equation for the parameters $M$, $a$, $\s$ and $R$,
which nonetheless factorizes and eventually leads to the quadratic
equation for $\s$,
\bea &&(R^2+2MR+4a^2)^2\s^2-M^2R^2(R+2M)^2 \nonumber\\
&&\hspace{2cm} +a^2(R^2-4M^2+4a^2) (R^2+4MR-4M^2+4a^2)=0,
\label{quade} \eea
with the positive root
\be \s=\sqrt{M^2-a^2+\frac{4M^2a^2(R^2-4M^2+4a^2)}
{(R^2+2MR+4a^2)^2}}, \label{sig} \ee
which coincides with the expression for $\s$ obtained in
Ref.~\cite{Cab}.

Taking into account (\ref{sig}), the constant quantity $d$ from
(\ref{cd}) assumes the form
\be d=\frac{2Ma(R^2-4M^2+4a^2)} {R^2+2MR+4a^2}, \label{d} \ee
and this is exactly the quantity $\delta$ from the paper
\cite{Cab}. The constant $K_0$ from (\ref{K0}) rewrites, with
account of (\ref{sig}) and (\ref{d}), as
\be K_0=\frac{4\s^2[(R^2+2MR+4a^2)^2-16M^2a^2]}
{M^2[(R+2M)^2+4a^2]}. \label{K0n} \ee
Mention that the above expression for $d$ can be also used for
writing $\s$ in a slightly simpler form
\be \s=\sqrt{M^2-a^2+d^2 (R^2-4M^2+4a^2)^{-1}}. \label{sig2} \ee

Therefore, the 3-parameter specialization of the MMR solution
describing two equal corotating Kerr black holes separated by a
strut is defined concisely by the formulas (\ref{mfn}),
(\ref{Rrn}) and (\ref{sig})-(\ref{K0n}). Apparently, our
expressions for the Ernst potential and for all metric functions
defining this subfamily are a good deal simpler than the ones
obtained in Ref.~\cite{Cab}.

On the horizons (the null hypersurfaces $\rho=0,
-\sigma<z-{\textstyle\frac{1}{2}}R<\sigma$ and $\rho=0,
-\sigma<z+{\textstyle\frac{1}{2}}R<\sigma$ --- two thick rods in
Fig.~1), the  black-hole constituents of this binary configuration
are expected to verify the well-known Smarr mass formula
\cite{Sma}
\be M=\frac{1}{4\pi}\kappa S+2\Omega J, \label{Sma} \ee
where $\kappa$ is the surface gravity, $S$ the area of the
horizon, $\Omega$ the horizon's angular velocity and $J$ the Komar
angular momentum of a black hole. Apparently, because of the
equatorial symmetry of the problem, the relation (\ref{Sma})
should be checked only for one of the constituents, say, for the
upper one. Since the black holes are corotating, their Komar
masses and angular momenta are both halves the respective total
values, $M_T$ and $J_T$, determined by (\ref{MJ}); hence, the mass
of each black hole is $M$, while the corresponding individual
angular momentum $J$ is given, as it follows from (\ref{MJ}) and
(\ref{d}), by the expression \cite{Cab}
\be J=\frac{Ma[(R+2M)^2+4a^2]}{R^2+2MR+4a^2}, \label{AM} \ee
and one can see that the inverse dependence $a(J)$ is defined by a
cubic equation.

For the calculation of the quantities $\kappa$, $\Omega$ and, the
following formulas should be used \cite{Tom}:
\be \kappa=\sqrt{-\omega_H^{-2}e^{-2\gamma_H}}, \quad
\Omega=\omega_H^{-1}, \quad S=4\pi\s\kappa^{-1}, \label{kap} \ee
where $\omega_H$ and $\gamma_H$ denote the values of the metric
functions $\omega$ and $\gamma$ on the horizon. The
straightforward calculations carried out for the upper black hole
yield the following expression for the horizon's angular velocity:
\be \Omega=\frac{(M-\sigma)(R^2+2MR+4a^2)}{2Ma[(R+2M)^2+4a^2]},
\label{Omega} \ee
while the quantities $S$ and $\kappa$ are defined by the formula
\cite{Cab}
\be S=\frac{4\pi\sigma}{\kappa}=\frac{8\pi
M[(R+2M)^2+4a^2][(R+2M)(M+\s)-4a^2]} {(R+2\sigma)(R^2+2MR+4a^2)}.
\label{S} \ee
Then it is easy to see that Smarr's relation (\ref{Sma}) is indeed
verified by virtue of (\ref{AM}), (\ref{Omega}) and (\ref{S}).

Let us òùö briefly comment on the possibility of the equilibrium
without a strut between two corotating Kerr sources. If we denote
by $\gamma_0$ the constant value of the metric function $\gamma$
on the strut, then the interaction force in our binary system can
be found by means of the formula ${\cal F}=(e^{-\gamma_0}-1)/4$
\cite{Isr,Wei}, thus yielding \cite{Cab}
\be {\cal F}=
\frac{M^2[(R+2M)^2-4a^2]}{(R^2-4M^2+4a^2)[(R+2M)^2+4a^2]}.
\label{F} \ee
This force becomes zero at infinite separation of the
constituents, and also when $|a|=(R+2M)/2$. In the latter case,
$\s$ becomes a pure imaginary quantity, which means that balance
at finite separation is only possible between two hyperextreme
Kerr sources; the value of the angular momentum leading to the
equilibrium is $|J|=M(R+2M)^2/(R+M)$, being characteristic of the
Dietz-Hoenselaers equilibrium configuration \cite{DHo}.

In order to have a somewhat better idea about the interaction
force in the generic case, it seems plausible to resort to some
approximations in (\ref{F}) for introducing the angular momentum
$J$ explicitly. Then we readily get from (\ref{AM}) and (\ref{F})
the following approximate formula for ${\cal F}$ as $R\to\infty$:
\be {\cal F}\simeq\frac{M^2}{R^2}+\frac{4M^4-12J^2}{R^4}
+\frac{80MJ^2}{R^5}+ O\left(\frac{1}{R^6}\right). \label{Fa} \ee
The form of the leading term in (\ref{Fa}) responsible for the
spin-spin interaction of corotating Kerr sources coincides with
the one already given by Dietz and Hoenselaers \cite{DHo} through
the analysis of two limiting cases of spinning particles in the
double-Kerr solution \cite{KNe}.

\section{Towards the description of two non-equal
Kerr black holes}

We will now outline a possible approach to treating the general
case of interacting non-equal Kerr black holes which is likely to
provide new information in the future about the spin-spin
repulsion force in binary systems of rotating bodies. This
approach consists in reparametrizing the general extended
2-soliton solution in the manner similar to the one already
applied to the equatorially symmetric case in the previous
sections. The starting point of such a procedure is the axis data
of the form
\be e(z)=\frac{z^2+a_1z+a_2} {z^2+b_1z+b_2}, \label{axisG} \ee
where $a_1$, $a_2$, $b_1$ and $b_2$ are four arbitrary complex
constants, together with the choice of the parameters $\a_n$ of
the extended soliton solution in the form slightly different from
(\ref{alfas}) (see Fig.~2),
\be \a_1=\frac{1}{2}R+\s_1, \quad \a_2=\frac{1}{2}R-\s_1, \quad
\a_3=-\frac{1}{2}R+\s_2, \quad \a_4=-\frac{1}{2}R-\s_2, \label{an}
\ee
$\s_1$ and $\s_2$ taking real or pure imaginary values (real
$\s$'s, as usual, define black holes, while pure imaginary $\s$'s
--- the hyperextreme objects). The elimination of the angular
momentum monopole parameter in (\ref{axisG}) with the aid of the
Fodor {\it et al.} method \cite{FHP} and fixing the origin of
coordinates by means of (\ref{an}) reduces the number of arbitrary
real parameters in the data (\ref{axisG}) to six overall, and the
procedure of introducing the parameters $\a_n$ into the axis data
described in Sec.~I then leads to the following expression for the
reparametrized data (\ref{axisG}):
\be e(z)=\frac{z^2-(M+ia)z+c+id} {z^2+(M-ia)z+g+ih},
\label{axisG2} \ee
where $M$ is the total mass, $a$ is the rotational parameter,
while the constant quantities $c$, $d$, $g$ and $h$ are defined as
follows:
\be c=s-\mu, \quad g=s+\mu, \quad d=\frac{1}{4a}(\tau+\delta),
\quad h=\frac{1}{4a}(\tau-\delta), \label{gh} \ee
with
\bea s&=&-\frac{1}{4}[R^2+2(\s_1^2+\s_2^2-M^2+a^2)], \nonumber\\
\delta&=&\epsilon\sqrt{\tau^2-\kappa}, \quad \epsilon=\pm1,
\nonumber\\ \tau&=&2R(\s_1^2-\s_2^2)-4M\mu, \nonumber\\
\kappa&=&a^2[16(\mu^2-s^2) +(R^2-4\s_1^2)(R^2-4\s_2^2)].
\label{dk} \eea
The six arbitrary real parameters involved in the axis data
(\ref{axisG2}) are hence $M$, $a$, $R$, $\s_1$, $\s_2$, $\mu$, and
one see that in the particular case $\mu=0$, $\s_1=\s_2=\s$ the
data (\ref{axisG2}) reduces to the equatorially symmetric data
(\ref{axis3}), albeit a formal redefinition $M\to2M$, $a\to2a$.

Using the general formulas of the paper \cite{MRu2}, we have
worked out the Ernst potential and the whole metric determined by
the axis data (\ref{axisG2}) in the following concise form:
\bea \E&=&\frac{A-B}{A+B}, \quad f=\frac{A\bar A-B\bar
B}{(A+B)(\bar A+\bar B)}, \quad e^{2\gamma}=\frac{A\bar A-B\bar
B}{{\cal K}_0R_+R_-r_+r_-}, \nonumber\\
\omega&=&2a-\frac{2{\rm Im}[G(\bar A+\bar B)]}{A\bar A-B\bar B},
\nonumber\\ A&=&[R^2-(\s_1+\s_2)^2](R_+-R_-)(r_+-r_-)
-4\s_1\s_2(R_+-r_-)(R_--r_+), \nonumber\\
B&=&2\s_1(R^2-\s_1^2+\s_2^2)(R_--R_+)
+2\s_2(R^2+\s_1^2-\s_2^2)(r_--r_+) \nonumber\\
&&+4R\s_1\s_2(R_++R_--r_+-r_-), \nonumber\\
G&=&-zB +\s_1(R^2-\s_1^2+\s_2^2)(R_--R_+)(r_++r_-+R) \nonumber\\
&&+\s_2(R^2+\s_1^2-\s_2^2)(r_--r_+)(R_++R_--R) \nonumber\\
&&-2\s_1\s_2\{2R[r_+r_--R_+R_--\s_1(r_--r_+)+\s_2(R_--R_+)]
\nonumber\\ &&+(\s_1^2-\s_2^2)(r_++r_--R_+-R_-)\}, \label{mfnG}
\eea
where the functions $R_\pm$ and $r_\pm$ are given by the
expressions
\bea R_\pm&=&\frac{\delta+2ia[M(\pm2\s_2+R)-2\mu]}
{\tau-ia[(\pm2\s_2+R)(\pm2\s_2+R+2ia)+4s]}
\sqrt{\rho^2+\left(z+\frac{1}{2}R\pm\s_2\right)^2}, \nonumber\\
r_\pm&=&\frac{\delta+2ia[M(\pm2\s_1-R)-2\mu]}
{\tau-ia[(\pm2\s_1-R)(\pm2\s_1-R+2ia)+4s]}
\sqrt{\rho^2+\left(z-\frac{1}{2}R\pm\s_1\right)^2}, \label{RrnG}
\eea
and the choice of the constant ${\cal K}_0$ in the formula for
$\gamma$ must preserve the asymptotic flatness of the solution.

In order to interpret the metric (\ref{mfnG}) as describing two
unequal Kerr black holes, it is necessary to solve the condition
$\omega=0$ on the part
$\{\rho=0,-\frac{1}{2}R+\s_2<z<\frac{1}{2}R-\s_1\}$ of the
$z$-axis. However, the bad thing is that, compared to the
equatorially symmetric case, the resulting explicit form of the
axis condition in the general case is extremely cumbersome, so
that really very powerful computers are needed for being able to
perform the required calculations in the analytical form. In spite
of that, the numerical analysis of the axis condition suggests
that the analytical treatment of the general case is still
possible in principle because this condition leads to the quartic
algebraic equation for the parameter $\mu$. We do not exclude that
some clever redefinitions of the parameters or fortunate
substitutions might cause the factorization of the axis condition
and the eventual resolution of the problem in a relatively compact
form on the basis of the metric (\ref{mfnG}). But the
accomplishment of this technically very complicated mission will
remain a task for the future.

\section{Conclusion}

Therefore, we have shown that the vacuum MMR solution is very fit
for the analytical description and study of the binary
configuration of corotating identical Kerr black holes, for which
we have worked out a concise representation that improves the one
obtained in Ref.~\cite{Cab}. We have restricted our consideration
exclusively to the case of the non-extreme constituents because
the extreme case of two equal or non-equal Kerr black holes is
described by a subclass of the well-known Kinnersley-Chitre
solution \cite{KCh} which was already identified and discussed in
our earlier work \cite{MRu3}.

We are convinced that in order to get a better insight into the
nature of the spin-spin interaction, future research should be
more concentrated on the configurations of non-equal spinning
bodies because, apparently, the cases of identical constituents
can be considered as degenerations of the respective generic cases
and hence could in principle hide some important information about
the real strength of the spin-spin repulsion or attraction. In
this respect, a good understanding of the systems of identical
spinning bodies is certainly necessary and brings us closer to the
description of more sophisticated binary configurations that
arise, for instance, within the framework of the general 2-soliton
spacetime (\ref{mfnG}).

\section*{Acknowledgments}
This work was partially supported by CONACYT of Mexico, and by
Project FIS2015-65140-P (MINECO/FEDER) of Spain.

\newpage

\begin{figure}[htb]
\centerline{\epsfysize=120mm\epsffile{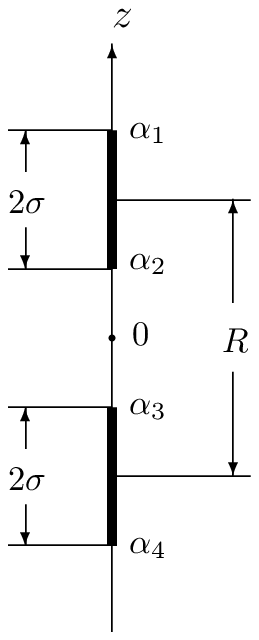}} \caption{Location
of two identical Kerr black holes on the symmetry axis:
$\a_4=-\a_1$, $\a_3=-\a_2$.}
\end{figure}

\begin{figure}[htb]
\centerline{\epsfysize=120mm\epsffile{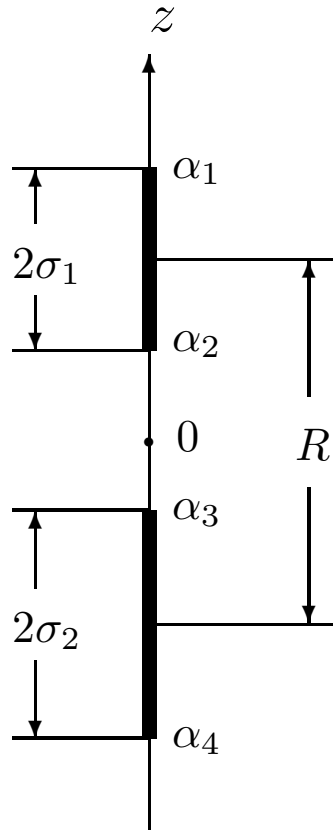}} \caption{Location
of two non-equal Kerr black holes on the symmetry axis.}
\end{figure}

\end{document}